\documentclass[aps,prc,twocolumn,superscriptaddress]{revtex4}
\usepackage{amsmath}
\usepackage{graphicx}

\usepackage{epsfig}

\def\b{{{ b}}}

\def\c{{{ c}}}
\def\d{{{ d}}}
\def\u{{{ u}}}
\def\q{{{ q}}}

\def\D{{{ D}}}

\def\T{{{ T}}}

\begin{document}
\title{Production and detection of doubly charmed tetraquarks}

\author{A. Del Fabbro}
\affiliation{INFN, Sezione di Trieste, I-34014 Trieste, Italy}
\author{D. Janc}
\affiliation{ Jo\v zef Stefan Institute, Jamova 39, SI-1001 Ljubljana, Slovenia}
\author{M. Rosina}
\affiliation{ Jo\v zef Stefan Institute, Jamova 39, SI-1001 Ljubljana, Slovenia}
\affiliation{ Faculty of Mathematics and Physics, 
University of Ljubljana, Slovenia}
\author{D. Treleani}
\affiliation{INFN, Sezione di Trieste, I-34014 Trieste, Italy}
\affiliation{ Universita di Trieste, Dipartimento di Fisica Teorica,
Strada Costiera 11, Miramare-Grignano}

\date{\today}


\begin{abstract}
The feasibility of tetraquark detection is studied. For the $cc\bar{u}\bar{d}$
tetraquark we show that in present (SELEX, Tevatron, RHIC) 
and future facilities (LHCb, ALICE) the production rate is promising 
and we propose some detectable decay channels.
\end{abstract}

\maketitle


\section{Introduction}

The purpose of this paper is to assess the possibility of detecting
certain tetraquarks in present and future facilities.
Among many possible tetraquarks, the double charm tetraquark
$T_{cc}=cc\bar{u}\bar{d}=DD^*$ with quantum numbers $IS^P=01^+$ is 
particularly interesting since it is very sensitive 
to the chosen effective interaction:
\begin{itemize}
\item It is very delicate, it is either weakly bound or slightly  unbound
with respect to the two-body hadronic decay $D+D^*$.
\item Its structure can be either predominantly ``molecular'' or 
predominantly ``atomic'' with consequences for the production and decay.
\end{itemize}

Double charm tetraquarks were intensively studied by many authors. 
Various approaches were applied, from lattice QCD and
chiral heavy quark effective theories
to nonrelativistic potential models. It was shown that, although 
the predictions of these theories agree in the barion and meson sector,
they give dramatically different results for tetraquarks.
For this reason, the double charm tetraquarks present an important
laboratory for discriminating between different hadronic models.

Moreover, our estimates for the production cross section of
such states gives us some hope that they can be experimentally
detected in near future.

The most important ingredient in the production of the
$\T_{\c\c}$ tetraquark is double charm production.
Experimental data for such events are very puzzling.
The production of prompt $J/\psi$ at $B$ factories as well as
the production of $\Xi_{\c\c}$ at SELEX are much larger than expected.
Therefore the comparison of  $\T_{\c\c}$ production with
$J/\psi c\bar{c}$ at B factories and the comparison 
$\T_{\c\c}$ production with $\Xi_{\c\c}$ at SELEX
could shed some new light on the mechanism responsible
for such large double charm production.

In Section 2 we presents our results for the $\T_{\c\c}$ production
at high energy colliders were we believe that the dominant
mechanism for the initial double charm production is double gluon fusion.
From experimental data we also estimate phenomenologically
the production of the $\T_{\c\c}$ tetraquark
at B factories and at SELEX.
The results of detailed four body calculations 
in nonrelativistic constituent quark model (Section 3)
encourage us to further investigate this state.
Since we found the $\T_{\c\c}$ tetraquark to be weakly bound, 
we propose (Section 4)
the branching ratio between hadronic and radiative decays as the most
promising mechanism for the detection of these states.

\section{ Production of double charm at various facilities }

The most promising mechanism for the production of the $\T_{\c\c}$ tetraquark 
is the formation of the $cc$ diquark followed by hadronisation 
into $\c\c\bar{\u}\bar{\d}$.
An alternative mechanism would exploit binding of
$\D$ and $\D^*$ mesons if when they are produced with small relative momenta.
One might expect that the latter mechanism could drastically enlarge 
the production rate if the dominant configuration is molecular.
Due to the very messy environment in hadron colliders, however, 
such a weakly bound system would too soon dissociate into free mesons 
by the interaction with surrounding partons of initial hadrons.

The first step is to create two $\c\bar{\c}$ pairs
with the $\c$ quarks close in the phase space
and in colour antisymmetric state, so that in the second step they
bind into the $\c\c$ diquark.
The binding energy of such a system is $\sim$ 200 MeV \cite{JR}.
In the third step the diquark gets dressed either with a light
$\u$ or $\d$ quark into a $\c\c\u$ or $\c\c\d$ baryon or with
a light $\bar{\u}\bar{\d}$ antidiquark
into the $\c\c\bar{\u}\bar{\d}$ tetraquark.
The probabilities for this two types of dressing can be
estimated using the analogy of a single heavy quark fragmentation.
The branching ratio of the $\b \to B$ and $b \to\Lambda_b$ production
at the Fermilab and at LEP
experiment is 0.9 and 0.1 \cite{dress},respectively, 
therefore we expect the same ratio in the hadronisation
$cc\to\Xi_{cc}$ and $cc\to\T_{\c\c}$, respectively.

The double charmed baryons were probably detected at SELEX \cite{sx}.
It was estimated that $40\%$ of the singly charmed baryons they see
result from the decay of doubly charmed baryons.
The most probable mechanism for the double charm production
at SELEX is production of the single $\c\bar{\c}$ pair in the processes
$gg\to \c\bar{\c}$  or $\q\bar{q}\to \c\bar{\c}$
while the second $\c\bar{\c}$ pair is created in the fragmentation
of the heavy quark $\c\to\c\c\bar{\c}$.
However, theoretically it is still unclear why the SELEX has such a large 
cross section for double charm production.
Since SELEX is a fixed target experiment the $\c\c$ diquark is most likely
to be produced with high lab momenta which might be helpful
in the detection as discussed in \cite{nussinov}.
But since SELEX found, with their cuts, only about fifty candidates for
double charmed baryons, the statistics for detecting double charmed
tetraquarks should be improved.

Next, we look at the
production and detection 
of the $\T_{\c\c}$ tetraquark in B-factories.
Since the total mass of four $D$ mesons is close to the
c.m. energy, the $c$ quarks created in this process have
small relative momenta which is very important in
$\T_{\c\c}$ production.
This feature also ensures a smaller number of additional pions created in
the $e^+e^-$ annihilation and thus a cleaner reconstruction of $\T_{\c\c}$.
Belle \cite{belle1}, \cite{belle2} has reported a measurement of
prompt $J/\psi$ production in $\mathrm{e}^+\mathrm{e}^-$
annihilation at $\sqrt{s}$ = 10.6 GeV
and found that the most of the observed $J/\psi$  production is due to the
double $\c\bar{\c}$ production
$$\sigma(\mathrm{e}^+\mathrm{e}^-\to J/\psi \c\bar{\c})/
\sigma(\mathrm{e}^+\mathrm{e}^-\to J/\psi X)=0.59^{+0.15}_{-0.13}\pm 0.12$$
which correspond to \cite{belle1}, \cite{belle2}
$$\sigma(\mathrm{e}^+\mathrm{e}^-\to J/\psi \c\bar{\c})=
0.87^{+0.21}_{-0.19}\pm 0.17\,\mathrm{pb}$$
or to about 2000 events from their 46.2 fb$^{-1}$ data sample.
The theoretical nonrelativistic QCD prediction for this process is
an order of magnitude smaller \cite{li6}, \cite{li8}, \cite{li9}, 
\cite{li10}, so this process is still not
well understood \cite{Ioffe}. But it is very likely that the
analogous mechanism  would also enlarge the cross section 
for the prompt production of
$\c\c$ diquark and thus the $\Xi_{\c\c}$ baryon and $\T_{\c\c}$ tetraquark.
The $\c\c$ diquark production cross section can then be estimated to be
$\sim$ 0.15 pb, which correspond to $\sim$ $10^4$ $\Xi_{\c\c}$
\cite{berezhnoy} and about $\sim$ $10^3$ $\T_{\c\c}$ per year.

We now present our calculation for the $\T_{\c\c}$ production
at high energy  colliders.
The two colliding nucleons in TeV machines can be considered 
as two packages of virtual gluons whose number is huge for
low Bjorken-$x$. Therefore we expect \cite{FT1}, \cite{FT2} 
that in these facilities the
dominant mechanism for double charm production would be a
double gluon-gluon fusion: $(g+g)+(g+g)\to (\c+\bar{\c})+(\c+\bar{\c})$.

The usual hard production mechanism is heavy quark production 
followed by fragmentation, however this mechanism does not include 
all the possible Feynman diagrams.
In ref. \cite{FT1}, \cite{FT2} it has been shown that at high energy collider
a significant rate of events with double heavy quark pairs is expected.
To compute all the fourth order $\alpha_s$ Feynman diagrams 
we have to consider  the two  different mechanisms leading to the same final
state:  the usual single parton scatterings  and the double parton
scatterings.
At high energy in single parton interactions the partonic sub-process is
dominated by the gluon-gluon fusion $ g g \rightarrow c \bar c c \bar c$ 
while the double parton interactions are dominated by the process
 $(g+g)+(g+g)\to (\c+\bar{\c})+(\c+\bar{\c})$
where the two distinct interactions occur in the same hadronic event.

\begin{figure}[here]
\resizebox{\linewidth}{\linewidth}{\includegraphics[width=\linewidth]{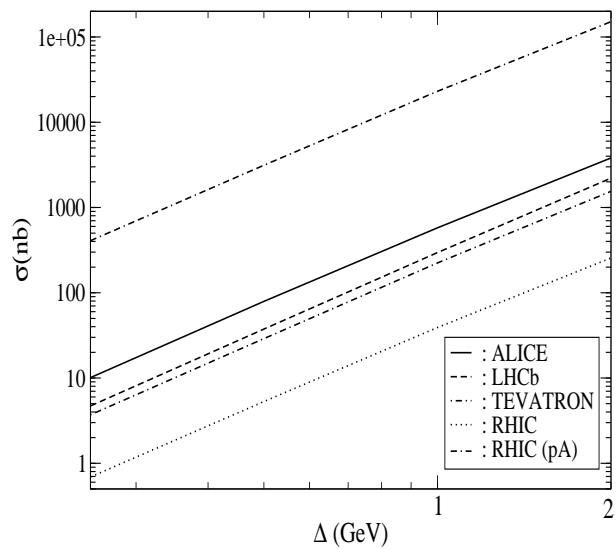}}
\caption {%
Production cross section of two $\c$ quarks in momentum space
$\Delta$ at LHC (LHCb and ALICE), at Tevatron
and at RHIC.
}
\label{sl0}
\end{figure}

We give an estimate of the production cross section at high energy in the
region of small transverse momenta where the multiple parton interactions 
provide the
leading contribution  to the cross section \cite{FT1}, \cite{FT2}.
We compute the production cross section of two $c$-quarks, $c_1, c_2$, very
close in momentum space $|p_{1j}-p_{2j}|<\Delta$,$\>$ $j =x,y,z$, 
as a function of $\Delta$.
We consider the heavy quark production in the kinematical range of the LHCb
$ (\sqrt s=14\,\mathrm{TeV},\;  1.8<\eta<4.9)$,  
 and for completeness for the ALICE
$ (\sqrt s=14 \,\mathrm{TeV}, \; |\eta|<0.9)$, 
Tevatron $ (\sqrt s=1.8 \,\mathrm{TeV}, \; |y|<1)$ and
RHIC
$ (\sqrt s=200 \,\mathrm{GeV}, \; |\eta|<1.6)$ experiments; in
the last case we calculate also the production cross sections in
proton-nucleus
interactions \cite{?}.  The results are shown in Fig \ref{sl0}
One can   notice that the cross section $d\sigma/d^3p$ at 
small $\Delta$ is almost uniform
and then it is approximately proportional to the momentum volume $\Delta^3$.

In the second step, the two $\c$ quarks join into a diquark.
We assume simultaneous production of two independent c quarks
with momenta $\vec{p}_1, \vec{p}_2$. Since they appear wherever within
the nucleon volume, we modulate their wave functions with a Gaussian profile
with the ``oscillator parameter'' $B=\sqrt{2/3}\sqrt{<r^2>}=0.69$ fm
corresponding to the nucleon rms radius
\begin{eqnarray*}
 & {\cal N}_B e^{-(\vec{r}_1-\vec{r}_a/2)^{\,2}/2B^2+{\rm i}\, 
 \vec{p}_1 \vec{r}_1}
  {\cal N}_B e^{-(\vec{r}_2+\vec{r}_a/2)^{\,2}/2B^2+{\rm i}\, \vec{p}_2 \vec{r}_2 }\\
  &\equiv
  {\cal N}_{(B/\sqrt{2})} e^{ -\vec{R}^2/2(B/\sqrt{2})^2+{\rm i}\,
    \vec{P} \vec{R}}
  {\cal N}_{(B\sqrt{2})} e^{ -(\vec{r}-\vec{r}_a)^{\,2}/2(B\sqrt{2})^2+{\rm i}\,
    \vec{p} \vec{r}}
\end{eqnarray*}
where the normalisation factor ${\cal N}_\beta=\pi^{-3/4}\beta^{-3/2}.$
Here $\vec{r}_a$ is the average distance between two nucleons in the target
nucleus for proton-nucleus experiment at RHIC and we use
the value $\vec{r}_a=$ fm or zero otherwise.

We make an impulse approximation that this two-quark
state is instantaneously transformed in any of the eigenstates of the
two-quark Hamiltonian. Then the amplitude of the
diquark formation $M$ is equal to the
overlap  between the two free quarks and the diquark
with the same centre-of-mass motion. By approximating
the diquark wave function  with a Gaussian with the
oscillator parameter $\beta= 0.41$ fm we get
\begin{eqnarray*}
M(p) &=& \int d^3 r \>
  {\cal N}_{B\sqrt{2}} e^{ (-(\vec{r}-\vec{r_a})^{\,2}/2(B\sqrt{2})^2-{\rm i}\,
     \vec{p} \vec{r})}
  {\cal N}_\beta e^{ (-\vec{r}^{\,2}/2\beta^2)} 
\end{eqnarray*}

For the production cross section we take into account that
$ d \sigma / d^3 p$ is practically constant and can be taken
out of the integral
\begin{eqnarray*}
\sigma &=& \frac{3}{9}\,\cdot\,\frac{3}{4}\int d^3 p\,\frac{d 
\sigma}{d^3 p} M^2(p)\\
   &\approx& \frac{1}{4}\frac{d \sigma}{d^3 p} \,
          \left(\frac{2\sqrt{\pi}\,\hbar}{\sqrt{2 B^2+\beta^2}}\right)^3 
          e^{-r_a^2/{2B^2}}
\end{eqnarray*}
where factors in front of the integral are due to the projection on the
colour and spin triplet states.
If we insert the values of $d\sigma/dp^3$ obtained from the 
Fig \ref{sl0}, we get  $\sigma\approx 27$ nb and $58$ nb
for LHCb and ALICE at LHC, $\sigma\approx 21$ nb at Tevatron and
$\sigma\approx 4$ nb and $63$ nb at RHIC for proton-proton and proton-nucleus
interaction, respectively.


The last step of the $\T_{\c\c}$ production is
dressing of the heavy diquark.
It either acquires one light quark to become
the doubly-heavy baryon ccu, ccd or ccs, or
two light antiquarks to become a tetraquark.
With this we neglect the possible dissociation of
the heavy diquark into a $DD$ pair so the results
are the upper estimate for the real $\T_{\c\c}$ production.
Assuming that the probability for dressing the $\c\c$ diquark into
the $\c\c\bar\u\bar\d$ tetraquark is 0.1 \cite{dress}, 
as pointed out at the beginning of the section, yields
the production rate of the dimeson 20900, 9700, 600 and 1 events/hour
for LHC at luminosity $10^{33}$ cm$^{-2}$s$^{-1}$, Tevatron 
at luminosity $8\cdot 10^{31}$ cm$^{-2}$s$^{-1}$ and RHIC
at $d-Au$ luminosity $0.2\cdot 10^{28}$ cm$^{-2}$s$^{-1}$, respectively.

\section{Structure of $\T_{\c\c}$}

The structure of the $T_{cc}$ tetraquark has been studied
in ref. \cite{fbx}. We summarize here those features which are 
particularly relevant for the detection.

There are two extreme spatial configurations of quarks in a tetraquark.
The first configuration which we call {\em atomic}
is similar to $\bar{\Lambda}_c$, with a compact
$cc$ diquark instead of $\bar{c}$,
around which the two light antiquarks are moving in a similar manner as in the
$\bar\Lambda_\c$ baryon. The second configuration which we call {\em molecular}
resembles deuteron, the two heavy
quarks are well separated and the two light antiquarks are bound to them
as if we had two almost free mesons. 
The atomic configuration is more likely to appear in strongly bound tetraquarks
while the molecular configuration can be expected in weakly bound systems. 

We present results using two different one-gluon exchange potentials.
The Bhaduri potential \cite{Bh} quite successfully describes the
spectroscopy of the meson, as well as baryon ground states. This is an
important condition since in the tetraquarks we have both quark-quark and
quark-antiquark interactions. The AL1 potential \cite{al1} slightly improves
the meson spectra by introducing a mass-dependent 
smearing of the colour-magnetic term.


We expand tetraquark wave function with Gaussinas of three sets of
Jaccobi coordinates. 
In this basis we were able to reconstruct the wave functions of
deeply bound tetraquarks 
as well as of two free mesons - the threshold state.
This is  important if one is  searching for weakly bound
tetraquarks with  molecular structure.
We found
that the $\T_{\c\c}$ is weakly bound for both the Bhaduri and  AL1 potential
in contrast to the results of calculations in harmonic oscillator 
basis \cite{SB} where asymptotic channel cannot be accommodated
as shown in Table \ref{tabela1}.

\begin{table}[here]
\caption{Column 1: type of potential, 
Column 2: lowest meson-meson threshold for a given potential in MeV,
Column 3: our results in MeV, 
Column 4: results in MeV of ref.\cite{SB}, 
Column 5: mean distance in fm between two heavy quarks
$\langle r_{\c\c}\rangle$. }
\label{tabela1}
\begin{tabular}{c c c c c}
\hline
  & threshold & our calc. & Ref.\cite{SB}& $\langle r_{\c\c} \rangle$  \\
\hline
Bhaduri &   3905.3 &  3904.7& 3931 & 2.4\\
AL1 &   3878.6 &      3875.9&      & 1.6\\
\hline
\end{tabular}
\end{table}

\begin{figure}[here]
\includegraphics[width=0.95\linewidth]{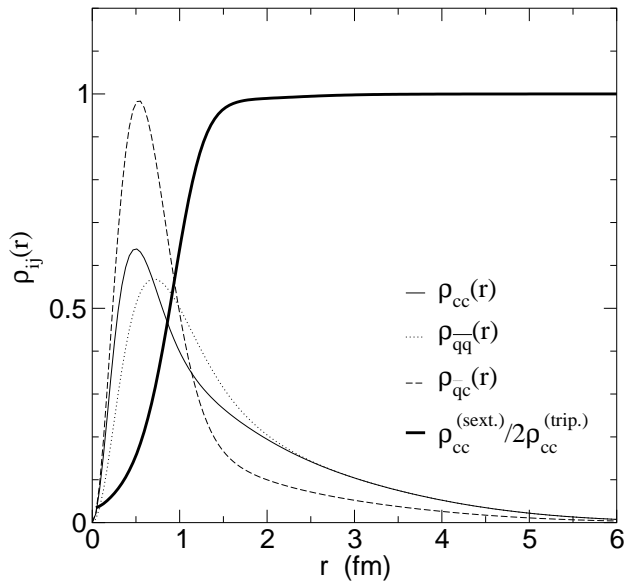}
\caption {%
Results for the AL1 potential.
Probability density of the two heavy quarks $\rho_{\c\c}$,
of the two light antiquarks $\rho_{\bar q\bar q}$ and of a light antiquark and
a heavy quark $\rho_{\bar q c}$ in $\T_{\c\c}$ as a function
of the interquark distance. The ratio of the projection on colour sextet and
colour triplet configurations is also shown.
}
\label{t1}
\end{figure}

In Fig. \ref{t1} we present the
probability densities $\rho_{ij}$ for finding (anti)quark $i$ and (anti)quark $j$
at the interquark distance $r_{ij}$ and the ratio of the projections on
colour sextet state $|6_{12}\bar 6_{34}\rangle_C$ and colour triplet state
$|\bar 3_{12}3_{34}\rangle_C$ where e.g.
$$\rho_{ij}^{\mathrm{\mathrm{(trip.)}}}(r)=
\langle \psi|\bar 3_{12}3_{34}\rangle_C \langle \bar
3_{12}3_{34}|_C \delta(r-r_{ij})|\psi\rangle.$$
Here particles 1 and 2 are the two heavy quarks
$\c$ and particles 3 and 4 the light antiquarks $\bar\u$ and $\bar\d$.
The wave function between heavy quarks
is broad and has an exponential tail $\sim \exp(-\kappa r)$ at large distances
where $\kappa =\sqrt{|E_{\mathrm{b}}|M_{red}}/\hbar c$, $E_{\mathrm{b}}$ is the binding energy
of the system and $M_{red}$ the reduced mass of the $\D$ and $\D^*$ mesons.
At small distances the dominant colour configuration is $\bar 3_{12}3_{34}$.
Here we have a diquark-antidiquark structure  and this region present about a third
of the total probability while for $r > 1$ fm sextet colour configuration
become larger.
The ratio of these two configurations stabilise at 2, since here we have
 a molecular structure of the two colour singlet mesons which has in diquark
 antidiquark basis 
$ |1_{13} 1_{24}\rangle =\sqrt{1/3}|\bar{3}_{12} 3_{34}\rangle+
\sqrt{2/3}|6_{12} \bar{6}_{34}\rangle$ colour decomposition, while
octet configuration 
$ |8_{13} 8_{24}\rangle =-\sqrt{2/3}|\bar{3}_{12} 3_{34}\rangle+
\sqrt{1/3}|6_{12} \bar{6}_{34}\rangle$ is negligible.


Now we show that additional weak three-body interaction can transform
the molecular structure of the $\T_{\c\c}$ tetraquark into atomic.
For the radial part we take the simplest possible radial dependence --
the smeared delta function of the coordinates of the three interacting
particles \cite{DDmol}. The colour factor in the two-body
Bhaduri or AL1 potential is proportional to the first (quadratic)
Casimir operator C$^{(1)}$; C$^{(1)}=\lambda\cdot\lambda$. It is then natural
that we
introduce in the three-body potential the second C$^{(2)}$ (cubic) Casimir
operator C$^{(2)}= d^{abc}\lambda_a\cdot\lambda_b\cdot\lambda_c$.
A deeper discussion of the properties that the colour dependent three-body
interaction must fulfil can be found in \cite{dma,dma2,s2}.

In the baryon sector the three-body interaction was used to better
reproduce the baryon ground state spectroscopy \cite{al1}. A colour structure is
there irrelevant since there is only one colour singlet state and thus the colour
factor is just a constant which can be included into the strength of the
potential. In tetraquarks the situation is different since there are
two colour singlet states: $\bar3_{12}3_{34}$ and  $6_{12}\bar6_{34}$
(or $1_{13}1_{24}$ and $8_{13}8_{24}$ after recoupling). The
three-body force operates differently on these two states and one can
anticipate that in the case of the weak binding it can
produce large changes in the structure of the tetraquark.
This cannot be otherwise produced simply by reparameterization
of the two-body potential, so the weakly bound tetraquarks
are a very important laboratory for studying the effect of such
an interaction.

\begin{figure}[h h h]
\includegraphics[width=0.83\linewidth]{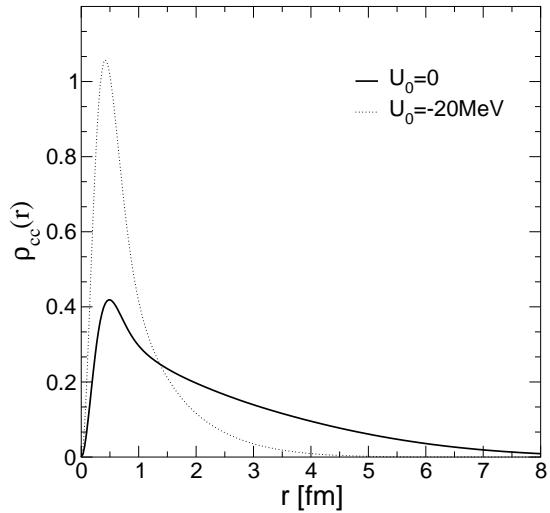}
\caption{%
Results for the Bhaduti potential.
Probability density between two c quarks $\rho_{cc}$ in the $\T_{\c\c}$
tetraquark as
a function of interquark distance
for three different values of the strength of the three-body potential.
}
\label{sl3body}
\end{figure}

A drastic change in the width of the probability density
can already be seen for strength $U_0$ = -20 MeV (Fig. \ref{sl3body}),
where the binding energy
of the tetraquark becomes - 14 MeV for Bhaduri potential. Here $\T_{\c\c}$
loses the molecular structure, the triplet-triplet colour configurations
become dominant and the $\T_{\c\c}$ tetraquark becomes similar to
$\T_{\b\b}$. In the baryon sector
such an interaction would merely lower the states by about $U_0$
so it would have no dramatic effect nor would it
spoil the fit to experimental data.
Since the predicted energies of ground state baryons for the
Bhaduri and AL1 potential are above the experimental values, this is
actually a desirable feature.

\section{Detection}

In order to identify a weakly bound $\T_{\c\c}$ tetraquark
we have to distinguish the pion or photon
emitted by the  $\D ^*$ meson bound inside the tetraquark
from the one resulting from free $\D^*$ meson decay.
We can exploit the fact that the phase
space for $\D ^*\to\D+\pi$ decay is very small.
This has a strong impact on the branching ratio
between radiative and hadronic decay. Since the  $\D ^*$ meson
inside the tetraquark with molecular structure is not significantly 
influenced by the other D meson in the tetraquark,
we expect that the partial width for the magnetic dipole
M1 transition would be very close to the width of the free meson
while the width for hadronic $\D ^*\to\D+\pi$ decay will decrease with stronger
binding and will become energetically forbidden below the $\D +\pi$ threshold.
The hadronic decay of the $\T_{\c\c}$ tetraquark is a three-body
decay which is commonly represented by the Dalitz plot.

\begin{figure}[h h h]
\includegraphics[width=0.85\linewidth]{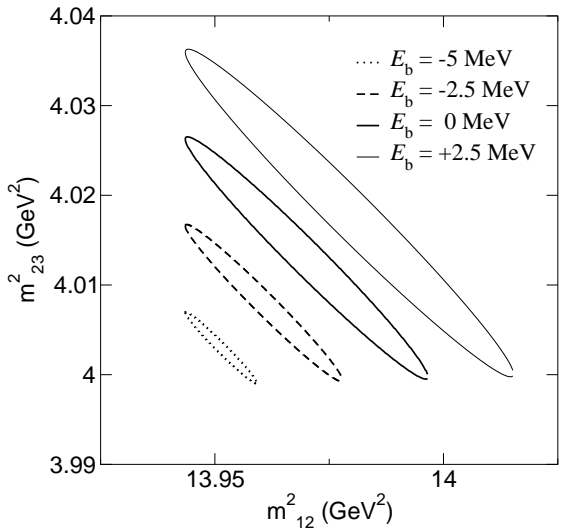}
\caption {%
Dalitz plot for four different values of the binding energy $E_{\mathrm{b}}$.
}
\label{sla}
\end{figure}

\begin{figure}[h h h]
\includegraphics[width=0.85\linewidth]{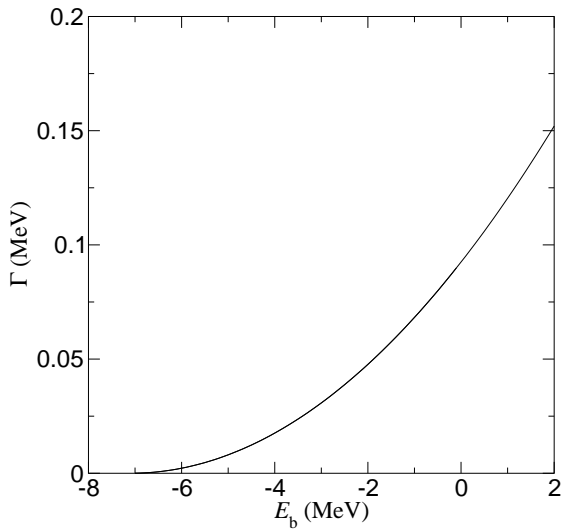}
\caption {%
Width of the $\T_{cc}$ tetraquark obtained with integration of the
 Dalitz plot $\int dm_{12}^2 dm_{23}^2/M^3$ where 
$M=m_D+m_{D^*}+E_{\mathrm{b}}$ is the mass of the tetraquark.
The width at $E_b=0$ is normalized to be equal to the width
of the free $\D^*$ meson decay.
}
\label{slb}
\end{figure}

If the $\T_{\c\c}$ tetraquark is below
the $\D+\D^*$ threshold but above the $\D+\D+\gamma$ and $\D+\D+\pi$,
as was the case in our nonrelativistic potential models,
the partial decay rate for the $\T_{\c\c}\to$D+D+$\pi$ is given by
\begin{equation}
d\Gamma=\frac{1}{(2\pi)^3}\frac{1}{32M^3}\overline{|{\cal M}|^2}
dm_{12}^2dm_{23}^2
\end{equation}
where particles 1 and 2 are two final $\D$ mesons and particle 3
is a $\pi$ emerging from the decaying tetraquark.
Here  $m_{12}^2=(p_D+p_D)^2$ and $m_{23}^2=(p_D+p_\pi)^2$
and M is the mass of the tetraquark.
Since the
total masses of the $\D ^*+D$ and $2\D+\pi$ are so close there is a strong
isospin violation in the decay which cannot be reproduced with
the Bhaduri or AL1 potential where the $\D ^*$ and the $\D$ isospin doublets
are degenerate.
We shall not try to modify the interaction to accommodate the dependence of the
decay on the isospin of the particles, but we shall rather
work with the experimental masses
taken from the PDG \cite{pdg}
where we see that $m_{\rm D^{*+}}-m_{\rm D^+}-m_{\pi^0}= 5.6 \pm 0.1\,{\rm MeV},\quad
m_{\rm D^{*0}}-m_{\rm D^0}-m_{\pi^0}=7.1 \pm 0.1\,{\rm MeV},\quad
m_{\rm D^{*+}}-m_{\rm D^0}-m_{\pi^+}= 5.87 \pm 0.02\,{\rm MeV}.$
The allowed region of integration over $dm_{12}^2$ and $dm_{23}^2$
for three different binding energies is plotted
in Fig. \ref{sla}. 
If we assume $\overline{|{\cal M}|^2}$ is constant, which is very
plausible in our case, the allowed
region will be uniformly populated with experimental events so that the
measured partial decay rate $\Gamma$ will be proportional to the kinematically
allowed area from Fig. \ref{sla}. 
This is shown in Fig. \ref{slb}, where we assumed that
for the molecular state, the width of the $\T_{\c\c}$ tetraquark with
zero binding energy would be the same as the width of the free
$\D^*$ meson.
\par
Let us now consider also the posibility that $\T_{\c\c}$ is
not a bound $\D\D^*$ state but a resonant state above the $D+D^*$
threshold.
Then if the
resonance is situated near the threshold,  there will be a significant
fraction of hadronic $\T_{\c\c}\to \D +\D +\pi$ decays beside the
$\T_{\c\c}\to \D +\D^*$ decay. This region of
positive binding energy is also presented in Fig. \ref{sla} and Fig. \ref{slb}.

In order to estimate the decay width
we make a comparison with charmonium.
The charmonium state $\psi (3770)$ has the width of $25.3\pm 2.9$ MeV and is
36 MeV above $\D\bar\D$ threshold,
which is also the dominant decay mode.
Let us assume, that the $\T_{\c\c}$ tetraquark resonant state, which would be
36 MeV above the $\D+\D^*$ threshold would have the same partial width for the
decay into $\D$ and $\D^*$ meson
The area of the integrated Dalitz plot for the binding energy 
$E_{\mathrm{b}}= +36$ MeV is then
37 times larger then at the threshold $E_{\mathrm{b}}= 0$.
Since the experimental width for $\D^*\to\D\pi$
is  $96\pm 4\pm 22$ keV \cite{pdg} we can estimate that the decay width for the
$\T_{\c\c}\to\D\pi\D$ three-body decay would be 
$\Gamma\sim 37\cdot 96 {\;\mathrm{keV}}=
3.6$ MeV. So we expect
about 15$\%$ direct $\T_{\c\c}\to\D+\D+\pi$ decays. The
Dalitz plot would not be uniformly populated but there 
will be a strong band where
$m_{23}=m_{D^*}$ reflecting the appearance of the  
$\T_{\c\c}\to\D\D^*\to\D\pi\D$
decay chain. In this estimation we have neglected the interference between
these two decay mechanism, since the width of the $\D^*$ is three orders of
magnitude smaller then the width of the tetraquark.


\section{Conclusion}

We have shown that the $T_{cc}$ tetraquark production is comparable
to double charm baryon production (possibly 10\%). Therefore they may be
seen in SELEX if statistics is improved. Similarly, it is comparable to
prompt $J/\psi c \bar{c}$ production which is reasonably abundant 
in $B$-factories. In high energy colliders we may expect an optimistic
number of events due to double $c\bar{c}$ production via  double
two-gluon fusion (see Sect.2). Therefore time has come to start the hunt!

Regarding the detection of the $\T_{\c\c}=\D\D^*$ tetraquark we propose
a nice opportunity -- the
very small phase space of the $\D^*\to\D\pi$ decay which is very
sensitive to the binding energy of $\D^*$ to $D$.
One possibility would be to measure the branching ratio between
the pionic and gamma decay of $D^*$.

{\bf{Acknowledgement.}}
This work was supported by the Ministry of Education, Science and Sport
of the Republic of Slovenia.


\begin{references}
\bibitem{JR} Janc D., Rosina M.:
Few-Body Systems {\bf 31}, 1 (2001)
\bibitem{nussinov} Gelman B. A., Nussinov S.:
Phys. Lett. {\bf B551}, 296 (2003)
\bibitem{li6} Cho P., Leibovich K.:
Phys. Rev. {\bf D54}, 6990 (1996).
\bibitem{li8} Yuan F., Qiao C.F., Chao K.T.:
Phys. Rev. {\bf D56}, 321 (1997); ibid, 1663 (1997).
\bibitem{li9} Baek S., Ko P., Lee J., Song H.S.:
J. Korean Phys. Soc. {\bf 33}, 97 (1998).
\bibitem{li10} Kiselev V.V. et al.:
Phys. Lett. {\bf B332}, 411 (1994).
\bibitem{Ioffe} Ioffe G.L., Kharzeev D.E.:
Phys. Rev. {\bf D69}, 0140016 (2004).
\bibitem{berezhnoy} Berezhnoy A.V., Likhoded A.K.:
Phys. Atom. Nucl. {\bf 67}, 757 (2004).
\bibitem{FT1} Del Fabbro A.,Treleani D.:
Phys. Rev. {\bf D61}, 077502 (2000).
\bibitem{FT2} Del Fabbro A.,Treleani D.:
Phys. Rev. {\bf D63}, 057901 (2001).
\bibitem{sx} Mattson M. et al. (SELEX Collaboration):
Phys. Rev. Lett. {\bf 89}, 112001 (2002)
\bibitem{dress} Affolder T. et al. (CDF Collaboration):
Phys. Rev. Lett. {\bf 84}, 1663 (2000)
\bibitem{belle1} Abe K. et al. (Belle Collaboration):
Phys. Rev. Lett. {\bf 88}, 052001 (2002)
\bibitem{belle2} Abe K. et al. (Belle Collaboration):
Phys. Rev. Lett. {\bf 89}, 142001 (2002)
\bibitem{nuss17} T\"ornqvist N. A.:
Phys. Rev. Lett. {\bf 67}, 556 (1991).
\bibitem{nuss19} T\"ornqvist N. A.:
Nuovo Cim. {\bf A107}, 2471 (1994).
\bibitem{nuss20} Manohar A. V., Wise M. B.:
Nucl.Phys. {\bf B399}, 17 (1993).
\bibitem{Bh} Bhaduri R. K., Cohler L. E., Nogami Y.: Nuovo Cim.
{\bf A65}, 376 (1981)
\bibitem{al1} Silvestre-Brac B.:
Few-Body Systems {\bf 20}, 1 (1996)
\bibitem{DDmol} Janc D., Rosina M.:
hep-ph/0405208
\bibitem{SB} Silvestre-Brac B., Semay C.:
Z. Phys. {\bf C57}, 273 (1993)
\bibitem{dma} Dmitrasinovic V.:
Phys. Lett. {\bf B499}, 135 (2001)
\bibitem{dma2} Dmitrasinovic V.:
Phys. Rev. {\bf D67}, 114007 (2003)
\bibitem{s2} Pepin S., Stancu Fl.:
Phys. Rew. {\bf D65}, 054032 (2002)
\bibitem{pdg} Hagiwara K. et al. (Particle Data Group):
Phys. Rev. {\bf D66} 010001 (2002)
\bibitem{fbx} Janc D., Rosina M.:
hep-ph/0405208
\end{references}
\end{document}